\documentclass[twocolumn,showpacs,amsmath,amssymb,prl]{revtex4}

\usepackage{amssymb,amsmath,bm}
\usepackage{graphicx}% Include figure files
\usepackage{dcolumn}% Align table columns on decimal point

\begin{document}

\title{Optical tornadoes in photonic crystals}
\author{Masaru Onoda$^{1}$ and Tetsuyuki Ochiai$^{2}$}
\affiliation{$^{1}$Department of Electrical and Electronic Engineering, Faculty of Engineering and Resource Science,
Akita University, 1-1 Tegata Gakuenmachi, Akita 010-8502, Japan} 
\affiliation{$^{2}$Quantum Dot Research Center, National Institute for
Materials Science (NIMS), 1-1 Namiki, Tsukuba 305-0044, Japan}
\date{\today}
\begin{abstract}
Based on an optical analogy of spintronics,
the generation of optical tornadoes is theoretically investigated
in two-dimensional photonic crystals without space-inversion symmetry.
We address its close relation to the Berry curvature in crystal momentum space, 
which represents the non-trivial geometric property of a Bloch state.
It is shown that the Berry curvature is easily controlled by
tuning two types of dielectric rods in a honeycomb photonic crystal.
Then, Bloch states with large Berry curvatures appear 
as optical tornadoes in real space.
The radiation force of such a configuration is analyzed,
and its possible application is also discussed.
\end{abstract}
\maketitle

Spintronics is usually discussed in terms of its aspect as magnetoelectronics, 
but it can be also interpreted as 
a technology to control electronic angular momentum via band(gap) engineering. 
The band engineering has been a key ingredient in electronics,
and nowadays also in optics \cite{Joannopoulos-PhC-book,Sakoda-PhC-book} 
and acoustics \cite{phononic-crystal-I,phononic-crystal-II},
where artificial materials with periodic structures
play the same role as solid crystals in electronics.
The band engineering in every system realizes
various types of functionalities 
by designing bandgaps and band dispersions of Bloch waves.
Electronics utilizes these factors
to control the electronic charge transport.
Then, spintronics attempts to control the electronic spin transport.
However, in contrast to the charge, 
each spin component is not necessarily conserved in the transport process
and intricately changes in some cases
due to the spin-orbit interaction.
However, at the same time, we can efficiently use this interaction to 
control the spin transport as demonstrated, for example, in the spin Hall effect
\cite{Kato,Wunderlich},
where applied electric field generates
the spin current perpendicular to the field.
Berry's geometric phase \cite{Berry} of a Bloch state is 
a key ingredient for the intrinsic mechanism of this effect \cite{MNZ,Sinova}.
More specifically, there appears 
non-zero Berry curvature in each Bloch band relevant to the effect.
In most cases, the spin-orbit interaction is essential
for the emergence of this non-trivial geometric property.

The spin-orbit interaction is a relativistic effect
and can be crucial not only in a specific crystal structure.
Photon is a relativistic particle of spin 1
and has the spin-orbit interaction in a extreme form
which is often represented as the transversality condition.
Consequently, electromagnetic waves in inhomogeneous media 
also show the non-trivial geometric property \cite{Chao, Liberman}.
For instance, the transverse shift of an optical beam reflected or refracted at an interface,
which is conventionally known as the Imbert-Fedorov effect
\cite{Fedorov,Imbert} in total internal reflection,
is naturally interpreted as the optical spin Hall effect
\cite{Onoda2004ohe,Bliokh,HK}.
The above mechanism of optical Hall effect was given a more general flavor
on the stage of a two-dimensional photonic crystal (PhC) which breaks 
the space-inversion symmetry (SIS) \cite{Onoda2004ohe,Onoda2006gao} and/or the time-reversal symmetry \cite{HR,Wang}.
When a parameter for the symmetry breaking is changed,
photonic bands show frequent crossing and repulsion with each other. 
There appear large Berry curvatures 
at nearly degenerate points in momentum space \cite{MN}.
It was also suggested that a wave-packet constructed from 
Bloch waves around such a state seems to involve
a kind of rotational motion \cite{Onoda2006gao}. 
However, any definite physical pictures in real space
have not yet been given to such Bloch waves.

In this paper, an optical analogy of spintronics in PhCs
is theoretically investigated.
That is, we attempt to utilize optical angular momentum via 
band engineering taking into account the above background information. 
For this purpose, we address how the Berry curvature in momentum space  
affects physical properties of Bloch waves in real space.  
We focus on the electromagnetic momentum density and the radiation force of a state with a large Berry curvature. 
It is shown that the Berry curvature of a photonic Bloch state in momentum space 
has a close relationship to the vortex structure of its momentum flow in real space.
This would open up a new way to realize an optical mixer which can selectively 
scramble and transport a specific kind of nanoparticles. 
It would serve for a broad range of research fields as well 
as the optical tweezers technology does.

\vspace{.5cm}
\noindent
\textit{Band engineering for the designing of Berry curvature ---}
The Berry curvature $\bm{\Omega}(\bm{k})$ of a photonic Bloch state
with a crystal momentum $\bm{k}$
stands for its geometric aspect \cite{Onoda2006gao}.
As well as the generality of band engineering
in several different systems, e.g., electronic, photonic and
phononic systems, the Berry curvature of a Bloch state
shares a common property in every system.
The most prominent one is its scaling behavior
around a nearly-degenerate point $\bm{k}_{0}$ \cite{MN}.
Especially for the typical case in a two-dimensional system, 
the scaling behavior is represented by 
\begin{eqnarray}
\Omega_z(\bm{k})
&\propto& \frac{v^2\Delta}{(v^2|\bm{k}-\bm{k}_0|^2+\Delta^2)^\frac{3}{2}},
\label{eq:Omega_z}
\end{eqnarray}
where $\Delta$ is the inter-band separation
and $v$ is a nominal velocity constant around $\bm{k}_{0}$.
This property enables us to control the Berry curvature
via the photonic band engineering.

Here we study this tunability in a honeycomb PhC \cite{Cassagne} without SIS.
The PhC consists of two (A and B) types of circular rods.
We can tune the difference between the two types of rods
through their dielectric constants $\varepsilon_{A(B)}$ or radii $r_{A(B)}$.
The dielectric constant of the background medium $\varepsilon_b$ 
is taken to be different from both of $\varepsilon_A$ and $\varepsilon_B$. 
When all the rods are identical,
the PhC recovers the SIS,
and its point group is $C_{6v}$. 
The point group at the K point ($G_\mathrm{K}$) is $C_{3v}$,
which has two one-dimensional irreducible representations
and one two-dimensional irreducible representation. 
Thus, the modes at the K point are either non-degenerate or doubly degenerate. 
When the SIS is broken,
the point group of the PhC and $G_\mathrm{K}$
becomes $C_{3v}$ and $C_3$, respectively,
and the doubly-degenerate modes are lifted.
For simplicity, we focus on the case $r_A=r_B$.
The degree of SIS breaking and 
the inter-band separation of the lifted modes at the K point
are controlled through $(\varepsilon_A-\varepsilon_B)$. 

\begin{figure}[h]
\includegraphics*[scale=0.3]{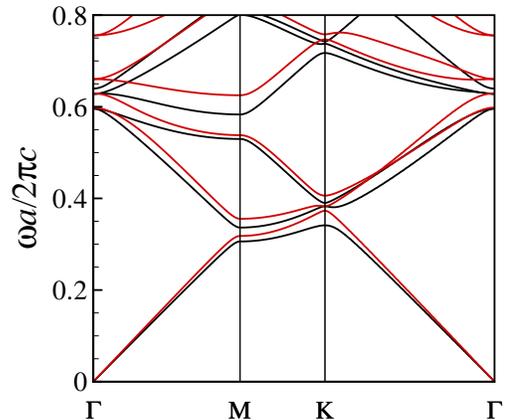}
\caption{\label{band} Photonic band structure of
 the honeycomb PhC with 
$\varepsilon_A=1.44, \varepsilon_B=1, \varepsilon_b=4,
r_A=r_B=0.2a$.
Black lines for the transverse magnetic (TM) mode
and red lines for the transverse electric (TE) mode.
}
\end{figure}

\begin{figure}[h]
\includegraphics*[scale=0.3]{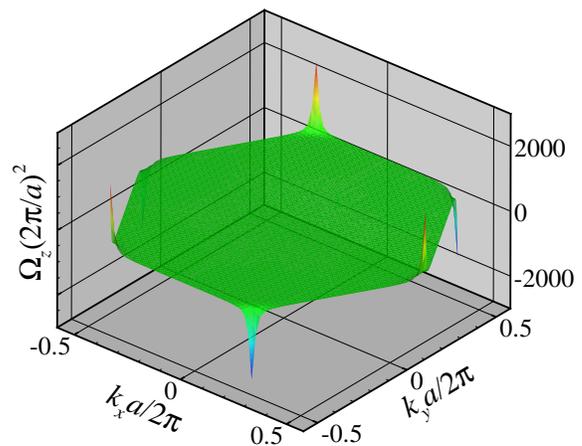}
\caption{\label{curvature} Berry curvature of the TE first band
in Fig.~\ref{band}.
}
\end{figure}

\vspace{.5cm}
\noindent
\textit{Berry curvature and optical tornado---}
Figure~\ref{band} shows
the photonic band structure of a honeycomb PhC
with the lattice constant $a$,
which is obtained by the photonic Korringa-Kohn-Rostoker method \cite{PhKKR}.
The black lines stand for the bands of transverse magnetic (TM) mode
and the red lines for those of transverse electric (TE) mode.
We can see various types of nearly degenerate points.
Here we focus on the nearly degenerate partners
of the TE first and second bands at the K point.
%%%%% Fig. 1 (band structure) %%%%%%%%%%%%%%%%%%%%
%%%%%%%%%%%%%%%%%%%%%%%%%%%%%%%%%%%%%%%%%%%%%%%%%%%%
The Berry curvature of the TE first band is shown in Fig. \ref{curvature}.  
As well as the pattern in Fig. \ref{curvature},
the Berry curvature of every band has the three-fold symmetry,
while every band dispersion exhibits six-fold symmetry.
For bands which are not globally degenerate in the Brillouin zone,
these symmetric patterns are understood by the following arguments.
With a symmetry operation $\mathcal{S}$ of the point group,
the energy dispersion $\omega(\bm{k})$ satisfies a simple relation
$\omega(\mathcal{S}^{-1}\bm{k})=\omega(\bm{k})$.
On the other hand, the Berry curvature $\bm{\Omega}(\bm{k})$
satisfies a relation for an axial vector field in momentum space,
just like a magnetic field in real space.
When the time-reversal symmetry is preserved, we can also tell
$\omega(-\bm{k})=\omega(\bm{k})$ and
$\bm{\Omega}(-\bm{k})=-\bm{\Omega}(\bm{k})$.
At a nearly degenerate point, the Berry curvatures of upper and lower bands
have the opposite sign with each other
because of the topological charge (Chern number) conservation \cite{Avron}.
Their signs interchange with each other
by reversing $\varepsilon_A$ and $\varepsilon_B$.
%%%%% Fig. 2 (berry curvature) %%%%%%%%%%%%%%%%%%%%%%%%
%%%%%%%%%%%%%%%%%%%%%%%%%%%%%%%%%%%%%%%%%%%%%%%%%%%%%%%%%%%

\begin{figure}[h]
\includegraphics*[scale=0.3]{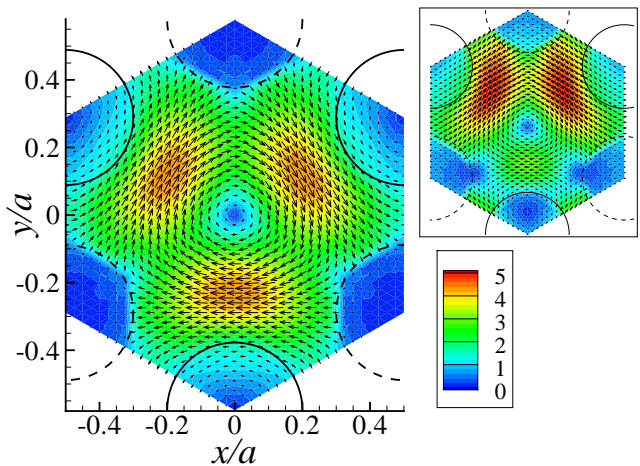}
\caption{\label{momentum} 
Minkowski momentum density of the Bloch state at the K point,
$(\frac{4\pi}{3a}, 0)$,
of the TE first band, see Fig.~\ref{band}.
Inset for the state at the point,
$(\frac{4\pi}{3a}-0.004\times\frac{2\pi}{a}, 0)$,
of the same band.
The circles of solid and dashed lines stand for
A-type and B-type rods, respectively.
In each panel, the magnitude is normalized 
by the average energy density/$c$.
Color contour stands for the absolute value
in the main panel and that multiplied by 4 in the inset.
}
\end{figure}

Next, let us check the real space configurations of states with large Berry curvatures.
Figure~\ref{momentum} shows the time-averaged Minkowski momentum density $\bm{D}\times\bm{B}$
of the Bloch state at the K point in the TE first band.
Only the configuration in a single Wigner-Seitz unit cell is shown.
The inset is for the state at a point slightly shifted from the K point in the same band.
In the present case, $\epsilon_{A(B)}<\epsilon_b$,
the momentum flow of the TE first band 
swirls mainly around the center of the unit cell.
However, in the case $\epsilon_{A(B)}>\epsilon_b$,
there can appear eddies mainly at the corners of the unit cell.
In any case, here we call such a Bloch state with vortex structure as an optical tornado.
This is because, just like a real tornado, 
the possible propagation direction of an optical tornado
can be perpendicular to its rotation vector,
as shown in the inset of Fig.~\ref{momentum}.
This point is a noticeable difference between
the optical tornado and
a Laguerre-Gauss beam with orbital angular momentum in free space.

There is a complementary property between the tornadoes of upper and lower bands
around a nearly degenerate point.
The orientation of a tornado around the K point is
clockwise for the TE first band (Fig.~\ref{momentum}), 
while that is counter-clockwise for the TE second band (not shown).
Here one may think that the tornadoes are specific to
the K point in the hexagonal Brillouin zone,
which is special in a sense that the point group is nontrivial.  
However, we can also find such tornadoes at a nearly-accidental-degenerate  
point of a square lattice PhC without SIS. 
The point group of this PhC is trivial, having no symmetry operation other than identity one.
%%%%% Fig. 3 (momentum flow) %%%%%%%%%%%%%%%%
%%%%%%%%%%%%%%%%%%%%%%%%%%%%%%%%%%%%%%%%%%%%%%%%%%%%%%%%%%%%%

%\vspace{.5cm}
\noindent
\textit{Radiation force induced by optical tornado---}
Finally, we consider the radiation force acting on a small particle in the PhC. 
Here we are assuming the situation in which the background medium is an ideal liquid with $\epsilon_b$.
When the particle size is sufficiently smaller than a typical wave length,
the particle behaves like an electric dipole $\bm{p}_e$ induced by the applied field of a Bloch state.
In the non-relativistic dynamics of a dipole,
the ponderomotive force is given in terms of the Lorentz force \cite{Gordon} as,
\begin{eqnarray}
\bm{F}&\cong&(\bm{p}_e\cdot\bm{\nabla})\bm{E}+\frac{d\bm{p}_e}{dt}\times\bm{B}.
\label{eq:ponderomotive}
\end{eqnarray}
The same effect should be also estimated in terms of
the Maxwell stress tensor $\tensor{T}$ containing both the contributions
of applied field and dipole radiation as
$\bm{F} = \int \tensor{T}\cdot d\bm{s}$,
where the integral is taken over a surface just outside the particle.
This leads to the estimation in Ref.~\cite{Ford} which is different from Eq.~(\ref{eq:ponderomotive}).
%\begin{eqnarray}
%\bm{F}&\cong&\frac{2}{3}(\bm{p}_e\cdot\bm{\nabla})\bm{E}+
%\frac{1}{3}p_{e,j}\bm{\nabla}E_j+\frac{2}{3}\frac{d\bm{p}_e}{dt}\times\bm{B}
%\end{eqnarray}
However, in both estimations, the time-averaged force takes the common form \cite{CN},
\begin{eqnarray}
\langle\bm{F}\rangle&\cong&
\frac{1}{2}\Re\left[\alpha_p^*(\omega) \tilde{E}_i^*\bm{\nabla}\tilde{E}_i\right]
\nonumber
\\
&=&\frac{\alpha_R(\omega)}{4}\bm{\nabla}|\bm{\tilde{E}}|^2+\frac{\alpha_I(\omega)}{2}
\Im\left[\tilde{E}_i^*\bm{\nabla}\tilde{E}_i\right]
\label{eq:force}
\end{eqnarray}
where $\bm{\tilde{E}}$ stands for the complex amplitude of a harmonic electric field
$\bm{E}(t)=\Re[\bm{\tilde{E}}e^{-i\omega t}]$,
$\alpha_p(\omega)=\alpha_R(\omega)+i\alpha_I(\omega)$ is the complex polarizability of the particle.
Readers can find the polarizability $\alpha_p(\omega)$
for a sphereical object of a radius $r_p$ with a complex permittivity $\epsilon_p(\omega)$
in Ref.~\cite{Draine},
\begin{eqnarray}
&&\alpha_p(\omega)\cong \alpha_p^{(nr)}(\omega)
\left[1-i\frac{(\epsilon_b\mu_b)^{\frac{3}{2}}\omega^3}{6\pi\epsilon_b}\alpha_p^{(nr)}(\omega)\right]^{-1},\\
&&\alpha_p^{(nr)}(\omega) \cong 4\pi\epsilon_b r_p^3\left[\frac{\epsilon_p(\omega)-\epsilon_b}{\epsilon_p(\omega)+2\epsilon_b}\right].
\end{eqnarray}
The last expression of Eq.~(\ref{eq:force}) suggests that the net force acting on the particle is divided into
the gradient force (the first term) and the scattering force (the second term).
Here it should be noted that 
the second term also includes the force via absorption
for particles with internal degrees of freedom, e.g., exciton resonance etc.
The scattering force is further divided into the momentum term and the rotational term as follows.
\begin{eqnarray}
\Im\left[\tilde{E}_i^*\bm{\nabla}\tilde{E}_i\right]
&=&\frac{\omega}{\epsilon_b}\Re\left[\tilde{\bm{D}}^*\times\tilde{\bm{B}}\right]
+\Im\left[(\tilde{\bm{E}}^*\cdot\bm{\nabla})\tilde{\bm{E}}\right]
\nonumber\\
&=&\frac{\omega}{\epsilon_b}\Re\left[\tilde{\bm{D}}^*\times\tilde{\bm{B}}\right]
-\frac{1}{2}\bm{\nabla}\times\Im\left[\tilde{\bm{E}}^*\times\tilde{\bm{E}}\right]
\label{eq:sc-force}
\end{eqnarray}
where we have used $\bm{\nabla}\cdot\tilde{\bm{E}}=0$.
It should be noted that the rotational term (the last term) vanishes 
for a simple plane wave and TM mode, but not necessarily for TE mode. 

For latter convenience, we extend the above argument
to the case of a line dipole with polarizability per unit length $\tilde{\alpha}_p(\omega)$.
The formula for the radiation force $\langle \bm{F}\rangle$
can be reinterpreted as the radiation force per unit length $\langle \tilde{\bm{F}}\rangle$
by replacing $\alpha_p(\omega)$ with $\tilde{\alpha}_p(\omega)$.
For a cylindrical object of radius $r_p$ with $\epsilon_p(\omega)$, 
the polarizability per unit length is estimated as
\begin{eqnarray}
&&\tilde{\alpha}_p(\omega)\cong \tilde{\alpha}_p^{(nr)}(\omega)
\left[1-i\frac{\mu_b\omega^2}{4l}\tilde{\alpha}_p^{(nr)}(\omega)\right]^{-1},
\label{eq:alpha-cylinder}
\\
&&\tilde{\alpha}_p^{(nr)}(\omega) \cong \left\{
\begin{array}{ll}
\pi r_p^2 [\epsilon_p(\omega)-\epsilon_b] & \mathrm{for\;TM\;mode} \\
2\pi\epsilon_b r_p^2\left[\frac{\epsilon_p(\omega)-\epsilon_b}{\epsilon_p(\omega)+\epsilon_b}\right] 
& \mathrm{for\;TE\;mode}
\label{eq:alpha-nr-cylinder}
\end{array}
\right.
,
\end{eqnarray}
where $l=1$ for TM mode and $l=2$ for TE mode.
The denominator of Eq.~(\ref{eq:alpha-cylinder}) represents the effect of radiation damping.

The above result tells us in which situation the characteristics of an optical tornado can be efficiently utilized,
while it should be regarded as an order estimation.
For the purpose of steady trapping, in which tornadoes  unfortunately lose their stage,
the gradient force must be dominant.
This is the situation required in the optical tweezers technology \cite{Ashkin},
and is realized under the condition $\alpha_R(\omega)\gg \alpha_I(\omega)$.
Therefore, the particle should not show resonant scattering nor absorption at the frequency of an applied harmonic field.
On the other hand, in order to utilize the vortex structure of a tornado, 
we must enhance the scattering force.
This situation is realized under the condition $\alpha_R(\omega)\ll \alpha_I(\omega)$.
However, from the above formulae, this is not so easy in Rayleigh regime,
i.e., $r_p/\lambda\ll 1$ where $\lambda$ is the relevant wavelength.
We need some resonance at the frequency of interest.
This is because a resonant scattering can help the efficient momentum transfer
from electromagnetic field to nanoparticles \cite{Iida}.
In order to investigate this problem more seriously,
we pursue a full electromagnetic analysis taking into account a resonance.

\begin{figure}[h]
\includegraphics*[scale=0.3]{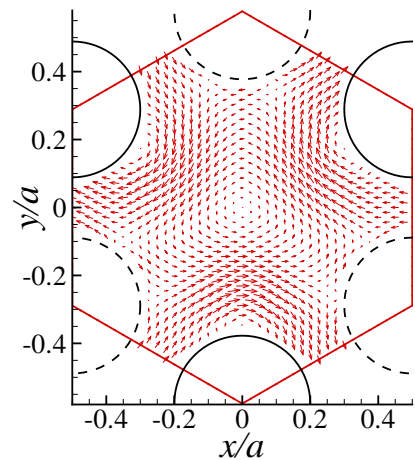}
\caption{\label{force} Radiation force per unit length
induced by the tornado at the K point
of the TE first band in Fig.~\ref{band}.
The probe is a thin metallic rod
with $\epsilon_p=-4.1$ and $r_p=0.02 a$.
The magnitude is normalized by the average energy density 
$\times$ $a$.
The arrow size is the actual size multiplied by 0.05.
}
\end{figure}

To this end, we evaluate the radiation force acting on a cylindrical probe via 
the formula $\langle\tilde{\bm{F}}\rangle=\int  \langle\tensor{T}\rangle\cdot d\bm{s}$, 
where the integral is taken over the cylindrical surface of a unit length.
The Maxwell stress tensor $\tensor{T}$ contains 
the contribution of the scattered wave from the cylindrical probe,
which is calculated by the t-matrix method \cite{Ochiai}.
The choice of the cylindrical shape for the probe is just for 
a technical simplification in the two-dimensional system.
In the analysis, a metallic property, i.e., $\epsilon_p<0$, 
is introduced to demonstrate an example of resonant scattering
due to a mechanism other than exciton excitation.
Although a naive expectation from Eqs.~(\ref{eq:alpha-cylinder}) and (\ref{eq:alpha-nr-cylinder})
suggests the resonance at $\epsilon_p=-\epsilon_b(=-4)$ for TE mode,
the full electromagnetic analysis for the finite size object with $r_p=0.02a$
predicts the slight shift of the resonant point, $\epsilon_p\cong -4.1$.
(For a metallic sphere, the resonance is expected around $\epsilon_p\cong -2\epsilon_b$.)
Figure~\ref{force} shows the results, and we can see that the radiation force
(per unit length) can take over a vortex structure from the optical tornado 
shown in Fig.~\ref{momentum} \cite{note-force}.
Its physical order of magnitude is about 2pN$\mu$m$^{-1}$
in the case with $a=1\mu$m and the radiation of 1mW$\mu$m$^{-2}$.
(Here we do not take into account the slow-light effect,
which enhances the magnitude of the force.)
The optical mixer via optical tornado
is sufficiently feasible when the resonant enhancement
of the momentum transfer is combined.
%%%%% Fig. 4 (radiation force) %%%%%%%%%%%%%%%%
%%%%%%%%%%%%%%%%%%%%%%%%%%%%%%%%%%%%%%%%%%%%%%%%%%%%%%%%%%%%%

%\vspace{.5cm}
\noindent
\textit{Summary---}
In summary, we have theoretically investigated 
the generation of optical tornadoes
and its close relation to the Berry curvature 
in a honeycomb photonic crystal without space-inversion symmetry.
It was demonstrated that the characteristics of an optical tornado 
is easily controlled via band engineering.
We also discussed the radiation force produced by the tornado acting on small objects.
An optical mixer via the tornado will be realized
by combining the resonant enhancement of efficient momentum transfer.
In other words, this functionality can be switched  selectively 
for a specific kind of nanoparticles.
This property would be useful for filtering or sorting
a mixture of different kinds of particles.

The authors thank K.~Sakoda, J.~Inoue and T.~Ishihara
for their insightful comments and useful information related to this work.


\begin{thebibliography}{99}
\bibitem{note-force}
We should note that the scattering force can 
be headed in the direction opposite to the Minkowski momentum. 
This is because the rotational term in Eq.~(\ref{eq:sc-force}) plays
a rather significant role.
Both the momentum and rotational terms exhibit a similar behaviour
except for their relative orientations.
\end{thebibliography}
\end{document}